# $B_2O_3$ glass former as a molecular matter revealed by heat capacity

Yan Zhuo Li, Ping Wen*, Wei Hua Wang

Institute of Physics, Chinese Academy of Sciences, Beijing 100190, P. R. China

Abstract

Heat capacity of $B_2O_3$ glass former in a wide temperature region is described well with the intrinsic motions for non-spherical $B_2O_3$ molecules, revealing that rather than a conventional network glass former, $B_2O_3$ is a typical molecular matter in which the transition from liquid to glass in the system corresponds to the frozen of translational motions for molecules.  The finding might provide an opportunity to understand the mysterious glass transition, as well as the intrinsic difference between solids and liquids.

PACS: 65.20.Jk, 65.60.+a

Boron trioxide $B_2O_3$ has attracted considerable attention because of its importance as a prototypical glass former and as an essential component of industrial glasses [1-13]. Despite these efforts, $B_2O_3$ is still puzzling case of study because it is uncertain whether $B_2O_3$ is a molecular system or not. Regarded usually as a typical network glass former with strong covalent interaction, $B_2O_3$ has been proposed to contain two elementary structure units, the planar $BO_3$ groups and the $B_3O_9$ boroxal rings [5,8]. The connectivity from corner-sharing planar $BO_3$ groups to a network entirely based on boroxal rings has been debated hotly [5]. Contrary to the above understanding, $B_2O_3$ has an obvious $\Delta C_P$ (around 40 $Jmol^{-1}K^{-1}$) at its low glass transition temperature $T_g$ (around 530 K), while the well-known typical network glass formers, such as $GeO_2$ and $SiO_2$, have a pretty small $\Delta C_P$ (around 1-3 $Jmol^{-1}K^{-1}$) at a higher $T_g$ [14-16]. Furthermore, a rapid decay in the instantaneous elastic modulus has been observed as temperature increases from $B_2O_3$ glassy region to its supercooled liquid region [17]. This softening has been found commonly in molecular glass forming liquids, rather than the typical network glass forming liquids [18,19]. Heat capacity built with the intrinsic motions and their carriers provides a powerful tool to understand the physical meaning of a macroscopic matter [20]. It has been confirmed successfully in simple solids and gases, where the vibration, the rotation and translational motion of elementary units are needed to be considered, respectively [20]. Recently a correlation between heat capacity and intrinsic motions of the basic particles for simple liquids has been founded [21]. The understanding to the physical nature of $B_2O_3$ becomes possible.

In this letter, the heat capacity of $B_2O_3$ is studied in a wide temperature range including the whole glassy region and most liquid region. A quantitative explanation to the heat capacity with the evolvements of the intrinsic molecular motions is set up, revealing that $B_2O_3$ glass former is a pure molecular system with a basic structural unit of $B_2O_3$ molecule, rather than the conventional network glass former. It is found that the excess heat capacity of $B_2O_3$ liquid relative to rigid solid is ascribed to the translational motions of $B_2O_3$ molecules. This clarification might be useful to understand the mysterious glass transition in simple glass formers.

$B_2O_3$ material with 99.999% (5N) purity was used for the heat capacity measurements. In order to reduce any water content, prior to the heat capacity measurements $B_2O_3$ was kept firstly for 24 hrs at 1173 K. Heat capacity in the low temperature region from 2 to 200 K was measured by Physical Property Measurement System using PPMS 6000 with an accuracy of about 2%. Heat capacity in temperature region from 170 to 860 K with an accuracy of about 5% was evaluated through the comparison with heat capacity of sapphire, and carried out by a Mettler Toledo DSC 1 thermal analyzer upon the heating or cooling at 20 K/min.

Heat capacity $C_P$ of $B_2O_3$ within a wide temperature region from 2 to 860 K is plotted in Fig. 1. According to the normal way, the heat capacity step $\Delta C_P$ is deduced to 39 Jmol$^{-1}$K$^{-1}$ at $T_g$ (533 K), and consistent with previous report [14,15]. It is necessary to highlight the difference between the liquid and glass, and $C_P$ of the liquid and glass around the glass transition in order to avoid the interference of the detailed kinetic process arisen from the glass transition [22]. $C_P$ around the glass transition considered here is simplified as the extrapolations of the experimental data at the temperatures higher than and lower than $T_g$, respectively. It is found that the $C_P$ tends to 0 Jmol$^{-1}$K$^{-1}$ as temperature is approaching to 0 K. With the increasing temperature the $C_P$ of the glass increases continuously. The temperature dependence of the $C_P$ in glass is much stronger than that in the liquid. Neglecting the detailed kinetic effect of the glass transition, we regard $C_P$ at $T_g$ jumps from the $C_P$ of the glass to that of the liquid upon the heating.

At $T_g$, the excess heat capacity $\Delta C_P$ of the liquid relative to the glass, the direct energy manifestation of the transition from a glass to corresponding liquid [23], offers the intrinsic difference between the solid and the liquid. According to the model of simple liquids proposed recently [21], a quantitative description to the $\Delta C_P$ for $B_2O_3$ glass former can be offered. The model is based on the physical picture as (a) liquids are composed up of tiny molecules which are posited in non-identical potential energy wells and (b) the motions of the molecule as a whole involve vibration, rotation (for non-spherical particles), and translational motion along the probing time axis and (c) among the molecular motions the translational motion, jumping out of one potential energy well into another well, is the key characterizing liquid state which never bear any shear. In this model, translational motion for molecule exist just in a liquid, rather than the corresponding solid within normally available probing time scale. Then the intrinsic difference between the liquid and the solid has a close relation to the translational contribution $C_P^{Tr}$ to $C_P$. The energy level of the molecular translational motion is given in classical physics as follow, $\varepsilon_{trans} = \sum_i (c_{xi} p_{xi}^2 + c_{yi} p_{yi}^2 + c_{zi} p_{zi}^2)$, where $p$ represents momentum, $x$, $y$, and $z$ denotes the Cartesian coordinates in space, and $i$ is the number of translational forms for the molecule considered. For simple molecules without radicals, the value of $i$ is equal to the number of rotational degree of freedom since in this situation only molecular rotation can be considered to affect the phase space determined by the translational motion. According to statistical mechanics, the value of $i$ is 3 for the non-spherical molecule. With the known mathematical treatment [24], the $C_P^{Tr}$ in the liquid containing one molar non-spherical molecule is $\frac{9}{2} R$ ($R$, gas constant). The $\Delta C_P$ at $T_g$ for $B_2O_3$ is 39 Jmole$^{-1}$K$^{-1}$, and is close to $\frac{9}{2} R$ (37.4 Jmole$^{-1}$K$^{-1}$) within the experimental error, indicating that it is appropriate in a theoretical consideration to regard $B_2O_3$ glass former as a molecular condensed matter in which the elementary structural unit, $B_2O_3$ molecule, is non-spherical in space.

The strength of the interactions between the molecules is usually characterized by Debye temperature $\Theta_D$. With the experimental data of $C_P$ at low temperatures close to 2 K (see Fig. 1), a linear relationship between $C_P$ and $T^3$ in the system is found in Fig. 2. $\Theta_D$ for $B_2O_3$ molecular vibration calculated by the equation [20], $C_P = \frac{12\pi^4 R}{5} (\frac{T}{\Theta_D})^3$, is about 117 K, meaning that the interaction between $B_2O_3$ molecules in the system is pretty weak. Consistent with the idea that $B_2O_3$ is a molecular system, Kasimer has reported that the heat capacity $C_P$ of $B_2O_3$ crystal was due mainly to vibrations of the molecules with weak interactions [25]. It is noted that the elementary structural units suggested by Kasimer is $B_4O_6$ molecule, but $B_2O_3$ molecule since $B_2O_3$ system was proposed to be distinguish with $Al_2O_3$ system.

In theory, $C_P$ of $B_2O_3$ glass can be described well by Debye model with the known $\Theta_D$ for $B_2O_3$ molecular vibration [20]. In practice, it is impossible to do that since the vibrational motions considered in Debye model are limited to those with low vibrational frequency. In the other words, Debye model is only applied well at temperatures lower than about $\Theta_D/50$, where only vibrations with low frequency can be activated thermally to contribute $C_P$ [26,27]. As temperature is higher than $\Theta_D/50$, more and more vibrations with higher frequency are activated to contribute to $C_P$, Debye model would lose its power to fit the experimental data. This has been confirmed by many studies [26-28]. Another possible reason lies on the fact that the vibrations for atoms inside the molecule would contribute $C_P$ at high temperature besides the molecular vibrations. That is, $C_P$ of rigid $B_2O_3$ glass contains at least two contributions. One is $C_P^{Vib,M}$ arisen from $B_2O_3$ molecular vibrations; the other is $C_P^{Vib,A}$ contributed from atomic vibrations inside the molecule.

Vibrational contribution to $C_P$ was formulated firstly by Einstein [29]. Regarding the vibrations for $j$ molecule as an independent system, the $C_P^{Vib,M}$ for a condensed matter containing $N$ identical molecules has the form as,

$$C_P^{Vib,M} = \int 3k_B \frac{\left(\frac{\Theta_E}{T}\right)^2 e^{\Theta_E/T}}{\left(e^{\Theta_E/T}-1\right)^2} g(\Theta_E)d\Theta_E,$$ where Einstein temperature $\Theta_E$ is equal to $\hbar\omega/k_B$ ($\hbar$ and $\omega$ are Plank constant and vibration frequency), and $g(\Theta_E)$ is the distribution of $\Theta_E$ corresponding to the vibration spectrum in the whole system. The value of 3 in the equation refers as to three dimensions for the molecular vibration. In order to calculate conveniently the above equation, Einstein regarded all vibrations as identical ones. This treatment is too rude to be true. In fact, whatever the complex $g(\Theta_E)$ for a real matter is, there is a generalized model to formulate $C_P^{Vib,M}$. Only if the equation, $\int g(\Theta_E)d\Theta_E = N$, is accepted at a given

temperature, the $C_P^{Vib}$ can be rewritten as, $C_P^{Vib,M} = 3Nk_B \dfrac{\left(\dfrac{\overline{\Theta_E}}{T}\right)^2 e^{\overline{\Theta_E}/T}}{\left(e^{\overline{\Theta_E}/T}-1\right)^2}$, where $\overline{\Theta_E}$ is also called as the characteristic Einstein temperature related to the distribution of $\Theta_E$ at a given temperature. The equation, $\int g(\Theta_E)d\Theta_E = N$, means the distribution of $\Theta_E$ is depended on temperature and becomes wider as temperature increases from 0 K since more and more vibrations are activated by the increasing thermal energy. As soon as all of vibrations with the possible $\Theta_E$ in a system are excited by thermal energy the distribution of $\Theta_E$ will be unvaried at high temperature because the possible $\Theta_E$ for a given system is fixed in theory. Correspondingly, $\overline{\Theta_E}$, as a function of temperature, has an general feature: It tends to increase monotonously up to a maximum with increasing temperature from 0 K, and be unvaried when temperature increases further. Similarly, the above modified Einstein's model can be applied to atomic vibration inside molecules since the vibration of one atom can be looked as an independent system. Since any atom in any molecule is posited with strong covalent bonds, it is appropriate to think the vibration for any atom with any covalent bond is in one dimension. So, $C_P^{Vib,A}$ related to the atomic vibrations for the matter containing $N$ particles is given form as, $C_P^{Vib,A} = f(n,n_l)Nk_B \dfrac{\left(\dfrac{\overline{\Theta_E}}{T}\right)^2 e^{\overline{\Theta_E}/T}}{\left(e^{\overline{\Theta_E}/T}-1\right)^2}$, where $f(n, n_l)$ is a integral number and depended on the number of atoms $n$ and bonds inside the particle $n_l$. On the basis of the electrochemical characteristics of B and O [25], non-spherical $B_2O_3$ molecule can be imaged as a symmetrical form as O=B−O−B=O in space. Therefore, the vibrational contribution to $C_P$ in $B_2O_3$ is written as,

$C_P^{Vib} = C_P^{Vib,M} + C_P^{Vib,A} = 3Nk_B \dfrac{\left(\overline{\Theta_E^M}/T\right)^2 e^{\overline{\Theta_E^M}/T}}{\left(e^{\overline{\Theta_E^M}/T}-1\right)^2} + 8Nk_B \dfrac{\left(\overline{\Theta_E^A}/T\right)^2 e^{\overline{\Theta_E^A}/T}}{\left(e^{\overline{\Theta_E^A}/T}-1\right)^2}$, where $\overline{\Theta_E^M}$ and

$\overline{\Theta_E^A}$, as functions of temperature, are characteristic Einstein temperatures for molecular vibration and atomic vibration inside B$_2$O$_3$ molecule.

Owning to that the intermolecular interaction is much weaker than intramolecular interaction, we suppose that the $C_P$ at low temperatures close to 0 K is contributed mainly from the of B$_2$O$_3$ molecular vibration, while the thermal effect of atomic vibrations appears only at temperature higher than 0 K. This consideration will bring a little error up, but the general tendency of $\overline{\Theta_E}$ can be accepted. The values of $\overline{\Theta_E^M}$ determined with the best fitting to the experimental data are plotted in Figure 3(b). It is found that $\overline{\Theta_E^M}$ increases monotonously from around 27 K to a maximum of 139 K as temperature increases from 2 to 49.5 K. When temperature increases further, $\overline{\Theta_E^M}$ fitting to experimental data starts to decrease. This decreasing in $\overline{\Theta_E^M}$ that has no clear physical meaning is arisen from the interference of the atomic vibrational contribution to the $C_P$. According to the above consideration, the value of $\overline{\Theta_E^M}$ is fixed as 139 K when temperature is higher than 49.5 K. The constant $\overline{\Theta_E^M}$ at temperature higher than 49.5 K is consistent with Brillion scattering experiment [17], where it is found the sound velocities approach to constants in the temperature region from 273 to 533 K (close to $T_g$). According to Debye model, Debye temperature $\Theta_D$ representing vibrations with low frequency is calculated to about 133 K. That the $\overline{\Theta_E^M}$ is higher than the $\Theta_D$ is reasonable because the former, rather than the latter, contains the vibrations with high frequency. The tendency of $\overline{\Theta_E^A}$ with temperature is similar as that of $\overline{\Theta_E^M}$ for molecular vibration. In Fig.3(b), $\overline{\Theta_E^A}$ increases up to a maximum (801 K) as temperature increases from 50 to 300K. Same as $\overline{\Theta_E^M}$, the value of $\overline{\Theta_E^A}$ is set to 801 K at temperatures higher than 300 K. Raman scattering study [30] revealing the basic feature of atomic vibrations has found that the temperature dependence of the main frequency shifts is very weak as temperature varies from 300 to 1300 K. So it is correct in a degree of approximation to set $\overline{\Theta_E^A}$ as a constant when temperature is higher than 300 K.

Figure 3(a) shows the $C_P$ at temperature less than 300 K can be described quantitatively by the vibrations in the system. However, as temperature is higher than 300 K, the vibrational contributions cannot follow the change of $C_P$ any more. This departure is ascribed to the missing contribution to $C_P$ in the system. In order to capture the missing contribution, residual heat capacity $C_P^R$, $C_P - C_P^{Vib,M} - C_P^{Vib,A}$, for the system is derived with the known $\overline{\Theta_E^M}$ and $\overline{\Theta_E^A}$ (see Fig. 3(b)). It should be emphasized that the $\overline{\Theta_E^M}$ will decrease in the liquid, where the softening exists. However, according to the modified Einstein's equation the decreasing $\overline{\Theta_E^M}$ does not affect the values of $C_P^{Vib,M}$ as temperature is higher than 300 K. Figure 4 exhibits the $C_P^R$ in the higher temperature region. It departs from zero at around 300 K, and tends to increase gradually up to around $3R/2$ as temperature increases up to $T_g$, finally becomes a constant (about $6R$) within the experimental error when temperature is higher than $T_g$. This confirms further that $B_2O_3$ glass former is a typical molecular condensed matter. Besides the vibrations of molecule as a whole and atoms inside molecules, rotations of molecules are also available at some temperatures. As a non-spherical molecule, owning to the rotation any excited $B_2O_3$ molecule contributes $3k_B/2$ to $C_P$ [20]. The number of the molecules excited thermally to rotate in the system increases gradually because of the continuous increase in $C_P^R$ as temperature increases from 300 to $T_g$. All of molecules are excited to rotate freely when temperature is approaching to $T_g$, where $C_P^R$ is close to $3R/2$. Above $T_g$ the constant $C_P^R$ is the exact superposition of the rotational contribution ($3R/2$) and the translational contribution ($9R/2$), revealing all of molecules are unlocked to rotate and move translationally in the liquid.

$B_2O_3$ glass former uncovered by $C_P$ as a molecular condensed matter is

meaningful to understand the softening phenomenon observed as temperature increases from the glassy region to its liquid region. It is not coincident that the softening takes place as soon as the translational motion for molecules is available. The occurring of any translational motions must make the local structures formed by molecules tend to be open and the interactions related to the local structures become weak. Correspondingly, instantaneous elastic modulus decreases and the softening takes place. Furthermore, it is intuitive that the averaged mobility of the translational motion increases as temperature increases. Then the possibility of the incompact local structures tends to be higher, and instantaneous elastic modulus decreases with the increasing temperature. In return, instantaneous elastic modulus is the characteristic parameter to the translational motion. Regarding translational motions for the molecules as the flow events in the liquid, one can predict that the dynamic behavior of the liquid is non-exponential with a distribution of energy barriers for the translational motions [31], and non-Arrhenius owning to the temperature dependence of the instantaneous elastic module [32]. The liquid dynamics behaviors arisen from the translational motions would be the further work in future.

In conclusion, with the clarification of heat capacity $B_2O_3$ glass former is demonstrated as a simple condensed matter composed up of non-spherical $B_2O_3$ molecules. The glass transition in the system is understood with the frozen of the translational motions for all of molecules in the liquid. The results provide an opportunity to uncover the mysterious properties of $B_2O_3$ glass former. Meanwhile, $B_2O_3$ glass former, as a typical molecular matter, could be an ideal system to clarify the long-standing issues, such as the glass transition and the physical nature of glasses and liquids.

**Acknowledgements**

The work has been supported financially from the Science Foundation of China (Grant Nrs: 51071170 and 11274353).

**Captions:**

Figure 1. The temperature dependence of heat capacity for B2O3 glass former within a wide temperature region.

Figure 2. Heat capacity $C_P$ at low temperature *versus* $T^3$ in order to drive the Debye temperature.

Figure 3. (a) The separation of heat capacity $C_P$ for $B_2O_3$ glass into three parts: $C_P^{Vib,M}$ arisen from $B_2O_3$ molecular vibrations, $C_P^{Vib,A}$ contributed from atomic vibrations inside the molecule and residual $C_P^R$, $C_P - C_P^{Vib,M} - C_P^{Vib,A}$, respectively. (b) The temperature dependence of the characteristic Einstein temperatures $\overline{\Theta_E^M}$ for molecular vibration and $\overline{\Theta_E^A}$ for atomic vibration inside $B_2O_3$ molecule; the dash lines are derived from the fitting to the experimental data (see details in text).

Figure 4. The temperature dependence of heat capacity $C_P$, $C_P^{Vib,M}$ arisen from $B_2O_3$ molecular vibrations, $C_P^{Vib,A}$ contributed from atomic vibrations and $C_P - C_P^{Vib,M} - C_P^{Vib,A}$ reduced by $R/2$ in higher temperature region.

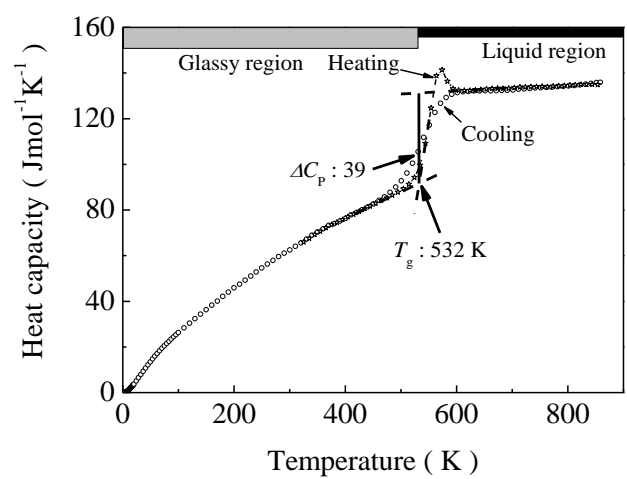

Figure 1.

Y. Z. Li *et. al.*,

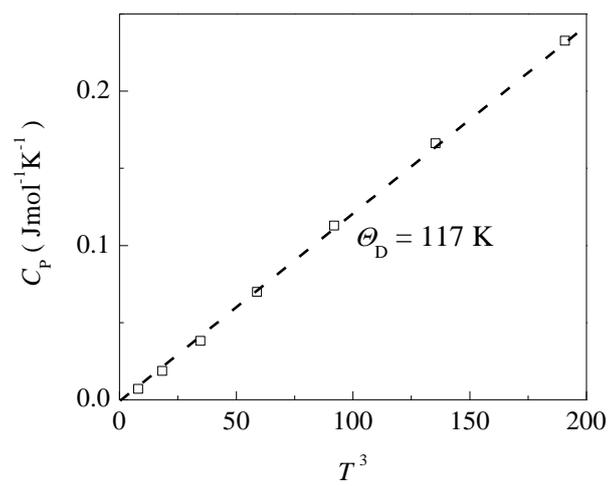

Figure 2.

Y. Z. Li *et. al.*,

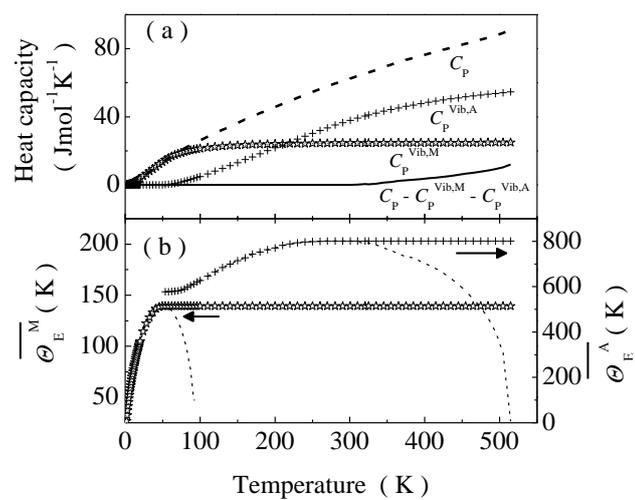

Figure 3.

Y. Z. Li *et. al.*,

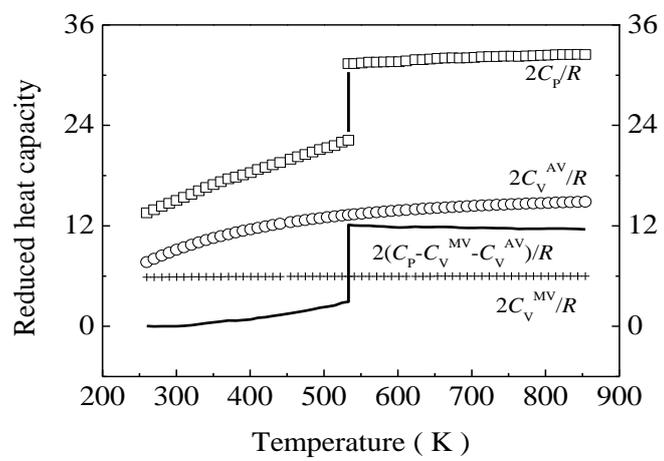

Figure 4.

Y. Z. Li *et. al.*,